\documentclass[a4paper,11pt]{article}
\usepackage{pos}

\usepackage{amsmath}
\usepackage{listings}
\usepackage{tikz}
\usetikzlibrary{positioning}
\usepackage{xcolor}
\usepackage{float}
\usepackage{graphicx}
\newcommand{\shuffle}{\mathbin{\raisebox{\depth}{\rotatebox[origin=c]{270}{$\exists$}}}}

\newcommand{\ourcode}{\texttt{HyperFORM}}
\newcommand{\form}{\texttt{FORM}}
\newcommand{\hyperint}{\texttt{HyperInt}}
\newcommand{\maple}{\texttt{Maple}}
\newcommand{\flint}{\texttt{FLINT}}
\newcommand{\ud}{{\rm d}}
\newcommand{\ue}{{\rm e}}
\newcommand{\ep}{\epsilon}
\newcommand{\EulerG}{\gamma_{\rm E}}
\newcommand{\CC}{{\mathbb C}}
\newcommand{\QQ}{{\mathbb Q}}

\definecolor{Mahogany}{cmyk}{0,0.85,0.87,0.35}
\definecolor{ForestGreen}{cmyk}{0.91,0,0.88,0.12}

\lstdefinelanguage{FORM}{
  alsoletter={.\#\_},
  morekeywords={
    Symbol,Symbols,Vector,Vectors,Index,Indices,Tensor,Tensors,
    Function,Functions,CFunction,CFunctions,Set,Dimension,AutoDeclare,
    Local,Global,Format,ModuleOption,
    symbol,symbols,vector,function,functions,cfunction,cfunctions,local,set},
  morekeywords=[2]{
    id,identify,repeat,endrepeat,also,multiply,Multiply,Print,print,
    Bracket,bracket,Keep,Collect,Drop,Skip,sum,Trace4,Tracen,Contract,
    Argument,EndArgument,Term,EndTerm,if,else,elseif,endif,while,Discard,
    SplitArg,FactArg,factarg,ArgToExtraSymbol,argtoextrasymbol,
    distrib_,prime_,replace_,gcd_,ToPolynomial,FromPolynomial},
  morekeywords=[3]{
    .sort,.end,.store,.global,.clear},
  morekeywords=[4]{
    \#include,\#define,\#redefine,\#call,\#do,\#enddo,
    \#procedure,\#endprocedure,\#if,\#else,\#endif,\#message,\#write,\#ifdef},
  morecomment=[f]{*},
  sensitive=true,
  morestring=[b]",
}

\lstdefinestyle{form}{
  language=FORM,
  basicstyle=\ttfamily\footnotesize,
  keywordstyle=\color{blue}\bfseries,        
  keywordstyle=[2]\color{purple},            
  keywordstyle=[3]\color{teal}\bfseries,     
  keywordstyle=[4]\color{Mahogany}\bfseries, 
  commentstyle=\color{ForestGreen}\itshape,  
  stringstyle=\color{orange},
  showstringspaces=false,
  breaklines=true,
  columns=fullflexible,
  keepspaces=true,
  literate={`}{{\char96}}1,
}

\title{\ourcode{} -- a \form{} package for parametric integration with hyperlogarithms}
\ShortTitle{\ourcode{}: parametric integration with hyperlogarithms in \form{}}

\author*[a,b]{Adam Kardos}
\author[c]{Sven-Olaf Moch}
\author[c]{Oliver Schnetz}

\affiliation[a]{Department of Experimental Physics, Institute of Physics,
  Faculty of Science and Technology, University of Debrecen,\\
  4010 Debrecen, PO Box 105, Hungary}
\affiliation[b]{Institute for Theoretical Physics, ELTE E\"otv\"os Lor\'and University,\\
  P\'azm\'any P\'eter 1/A, H--1117 Budapest, Hungary}
\affiliation[c]{II.\ Institut f\"ur Theoretische Physik, Universit\"at Hamburg,\\
  Luruper Chaussee 149, D--22761 Hamburg, Germany}

\emailAdd{kardos.adam@science.unideb.hu}
\emailAdd{sven-olaf.moch@desy.de}
\emailAdd{schnetz@mi.uni-erlangen.de}

\abstract{%
\ourcode{} brings the parametric integration of hyperlogarithms, weighted by
rational prefactors, into the symbolic-manipulation system \form{}. It ports
the capabilities of Erik Panzer's \maple{} package \hyperint{}, capitalizing on
\form{}'s speed with bulky algebraic input and on its ability to spread a single
calculation across many processor cores. We keep the description of the method
brief and concentrate instead on how the package is organized and driven: a
fully self-contained program for the three-loop zigzag period serves as a
worked illustration, and timing measurements for zigzags through six loops gauge
its present reach. \ourcode{} is released openly and applies to a broad class of
problems, the evaluation of Feynman integrals prominently among them.}

\FullConference{%
  Loops and Legs in Quantum Field Theory (LL2026)\\
  12--17 April, 2026\\
  Bayreuth, Germany\\}

\begin{document}
\maketitle
\raggedbottom

\section{Introduction}
Among the many techniques developed for high-order perturbative
calculations~\cite{Weinzierl:2022eaz}, those that return analytic answers hold a
special place: beyond a number, they lay bare the algebraic and number-theoretic
structures that organize the perturbative series. Parametric integration of
hyperlogarithms is one such technique. Embedded in the theory of graphical
functions, it has pushed the renormalization of scalar field theories to
unprecedented orders: six loops in $\phi^3$~\cite{Schnetz:2025wtu} and seven
loops in $\phi^4$~\cite{Schnetz:2022nsc}.

The approach is due to Francis Brown, who introduced it as a general-purpose
route to the analytic evaluation of a wide class of
integrals~\cite{Brown:2008um}. Its applicability rests on a single requirement:
each successive integration must land back in the space of hyperlogarithms.
Integrals that satisfy this condition are termed \emph{linearly reducible}. Two
independent realizations of Brown's algorithm were written in \maple{}, one by
Bogner~\cite{Bogner:2015nda} and one by Panzer~\cite{Panzer:2014caa}; the
latter, distributed as \hyperint{}, became the de-facto standard for the
linearly reducible Feynman integrals within its reach, up to intermediate loop
orders.

The method is, however, computationally heavy: intermediate expressions swell
quickly, an intrinsic feature of the algorithm, and in \hyperint{} this is
compounded by the host system, where \maple{} manages large expressions but is
not the most efficient choice, and by the program's limited development after
release. Our remedy is a faster engine built on
\form{}~\cite{Tentyukov:2007mu,Ruijl:2017cxj}, chosen because it processes large
algebraic expressions rapidly, is freely available and open, and parallelizes
efficiently across the many cores of present-day servers. As a first milestone we
have reimplemented most of \hyperint{}'s functionality in \form{}, released as
\ourcode{}~\cite{kardos_2025_17706909}; since no new mathematics enters here, the
present contribution focuses on what the package offers and how it is used, with
the underlying theory left to Refs.~\cite{Brown:2008um,Panzer:2014gra} and a
detailed account in the full paper~\cite{kardos_2025_17706909}.

\section{Hyperlogarithms and parametric integration}
\label{sec:method}
At the heart of the method sit the hyperlogarithms, also known as Goncharov
polylogarithms~\cite{Goncharov:1998kja}. A convenient way to characterize them is
through the recursion
\begin{equation}\label{eq:Ldef}
  \partial_z L(a_n,\ldots,a_1;z) = \frac{L(a_{n-1},\ldots,a_1;z)}{z-a_n}\,,
  \qquad L(a_n,\ldots,a_1;0) = 0\,,
\end{equation}
with $a_1,\ldots,a_n,z\in\CC$ and $\log(0)=0$ taken as the convention for their
evaluation. Equivalently they are Chen iterated integrals~\cite{Chen:1977oja},
and they close under the shuffle product
$L(u;z)\,L(v;z) = L(u\shuffle v;z)$, the sum $u\shuffle v$ running over every
shuffle of the words $u$ and $v$. When each letter $a_i$ is $0$ or $-1$ (the
hyperlogarithm analogue of the $\{0,1\}$ alphabet of harmonic polylogarithms),
the regularized value at $\infty$ is a multiple zeta value (MZV)
$\zeta(n_r,\ldots,n_1)$. At the
modest weights relevant here, these MZVs can simply be looked up:
\ourcode{} carries a reduction to a $\QQ$ basis taken from the
\emph{Multiple Zeta Value Data Mine}~\cite{Blumlein:2009cf} bundled with
\form{}.

The prime application is to Feynman integrals themselves. Introducing Feynman
parameters, an $L$-loop integral with $N$ propagators can be written as
\begin{equation}\label{eq:feynpar}
  I = \frac{\ue^{L\ep\EulerG}\,\Gamma\!\left(\nu-\tfrac{Ld}{2}\right)}
           {\prod_{i=1}^{N}\Gamma(\nu_i)}
  \left(\prod_{i=1}^{N}\int_0^\infty\!\ud x_i\right)
  \delta\!\left(1-\sum_{i=1}^N x_i\right)
  \left(\prod_{i=1}^{N}x_i^{\nu_i-1}\right)
  \frac{\mathcal{U}^{\,\nu-\frac{(L+1)d}{2}}}{\mathcal{F}^{\,\nu-\frac{Ld}{2}}}\,,
\end{equation}
with $\mathcal{U}$ and $\mathcal{F}$ the first and second Symanzik polynomials,
$L$ the number of loops and $d=4-2\ep$ the spacetime dimension.
Hyperlogarithmic integration exploits two structural properties: each $x_i$
enters $\mathcal{U}$ only to the first power, and the same holds for $\mathcal{F}$
in the absence of internal masses. The projectivity encoded in the $\delta$
function is eliminated by the Chen--Wu gauge choice, by which any single
parameter may be frozen to $1$ while the rest are integrated over
$[0,\infty)$. Splitting the (linear) polynomials in the first parameter into
partial fractions yields logarithms; on repeating the step the arguments of
those logarithms come to depend on the following parameter, and this is precisely
how hyperlogarithms arise.

One subtlety deserves emphasis. The original integral is finite, yet linearity
makes the algorithm break it into separate pieces, and an individual piece may
blow up like $\log(0)^k$ or $\log(\infty)^k$. Such terms are tamed by
\emph{regularized limits}~\cite{Panzer:2014gra}: $\log 0$ and $\log\infty$ are
carried along as formal constants which, for a convergent integral, cancel out
of the answer. \ourcode{} tracks them throughout and verifies, by default, that
the cancellation indeed takes place.

In practice, then, the integrals \ourcode{} targets are those whose
$\ep$-expansion is to be obtained order by order, of the generic shape
\begin{equation}\label{eq:integral-def}
  I(\ep) = \int_{0}^{\infty}\!\ud x_1\cdots\int_{0}^{\infty}\!\ud x_N\,
  \prod_{j=1}^{M} f_j^{\,n_j+m_j\ep}(x_1,\ldots,x_N)\,,
  \qquad n_j\in\mathbb{Z},\ m_j\in\QQ\,,
\end{equation}
the $f_j$ being rational in the integration variables. Whether or not such an
integral carries poles in $\ep$, the divergent case must first be cast into an
expandable form by the regularization of Ref.~\cite{Panzer:2014gra}, performed
ahead of the Laurent expansion.

\section{The HyperFORM package}
\label{sec:package}
The distribution~\cite{kardos_2025_17706909} comprises three \form{} headers.
The procedures themselves reside in \texttt{hyperform.h}; the companion
\texttt{declare-hyperform.h} gathers every symbol, function and
prepro\-cessor-variable declaration; and \texttt{mzvlow.h} stores the MZV reduction
table through weight ten~\cite{Blumlein:2009cf}, which suffices to carry a
calculation to the sixth loop order. Every identifier the package introduces is
namespaced, with \texttt{HYP} for symbols and functions and \texttt{Hyp} for
procedures, so that user-chosen names can never collide with internal ones.
Greatest-common-divisor and related polynomial manipulations are delegated to
\flint{}~\cite{flint,7891956} through \form{}'s built-in interface.

The package handles both finite and divergent integrals in $\ep$: it
expands non-integer powers, integrates rational functions as well as
hyperlogarithms, evaluates regularized limits, performs series expansions,
rewrites limits in a fibration basis, and reduces the outcome to MZVs.

\paragraph{Workflow.}
A calculation is a short \form{} program. The user includes the package, defines
the integrand as a local expression in terms of numerator/denominator
polynomials with $\ep$-dependent exponents, and then calls a fixed sequence of
procedures:
\begin{itemize}
  \item \texttt{HypParseInputExpr} translates the user integrand to the
        internal representation;
  \item \texttt{HypAutoRegularize} detects and regularizes $\ep$-divergences
        (recursive and automatic; leaves $\ep$-finite integrands unchanged);
  \item \texttt{HypEpExpand} expands in $\ep$ up to the order set by
        \texttt{HYPMAXEP};
  \item \texttt{HypApplyChenWu} fixes one integration variable to $1$ for
        projective (Feynman) integrals;
  \item \texttt{HypIntegrationStep} integrates over one variable, called in a
        loop over a chosen integration sequence;
  \item \texttt{HypFinalizeResult} produces a human-readable result (MZV
        shorthand, $\ep$-expansion to the requested order).
\end{itemize}
The choice of integration sequence matters: for more complex integrals some
sequences are linearly reducible while others are not, and different reducible
sequences can have very different runtimes.

\paragraph{Test-driven development.}
\ourcode{} follows a strictly modular design: if a routine needs more than one
sentence to describe, it is broken up. Each of the roughly~$90$ procedures comes
with its own unit test (built on the \texttt{check.rb} script shipped with
\form{}), exercising several cases. Development follows the red-green-refactor
cycle, so that the algorithms, which are not final, can be updated safely,
with regressions caught immediately.

\begin{figure}[H]
  \centering
  \includegraphics[width=0.9\textwidth]{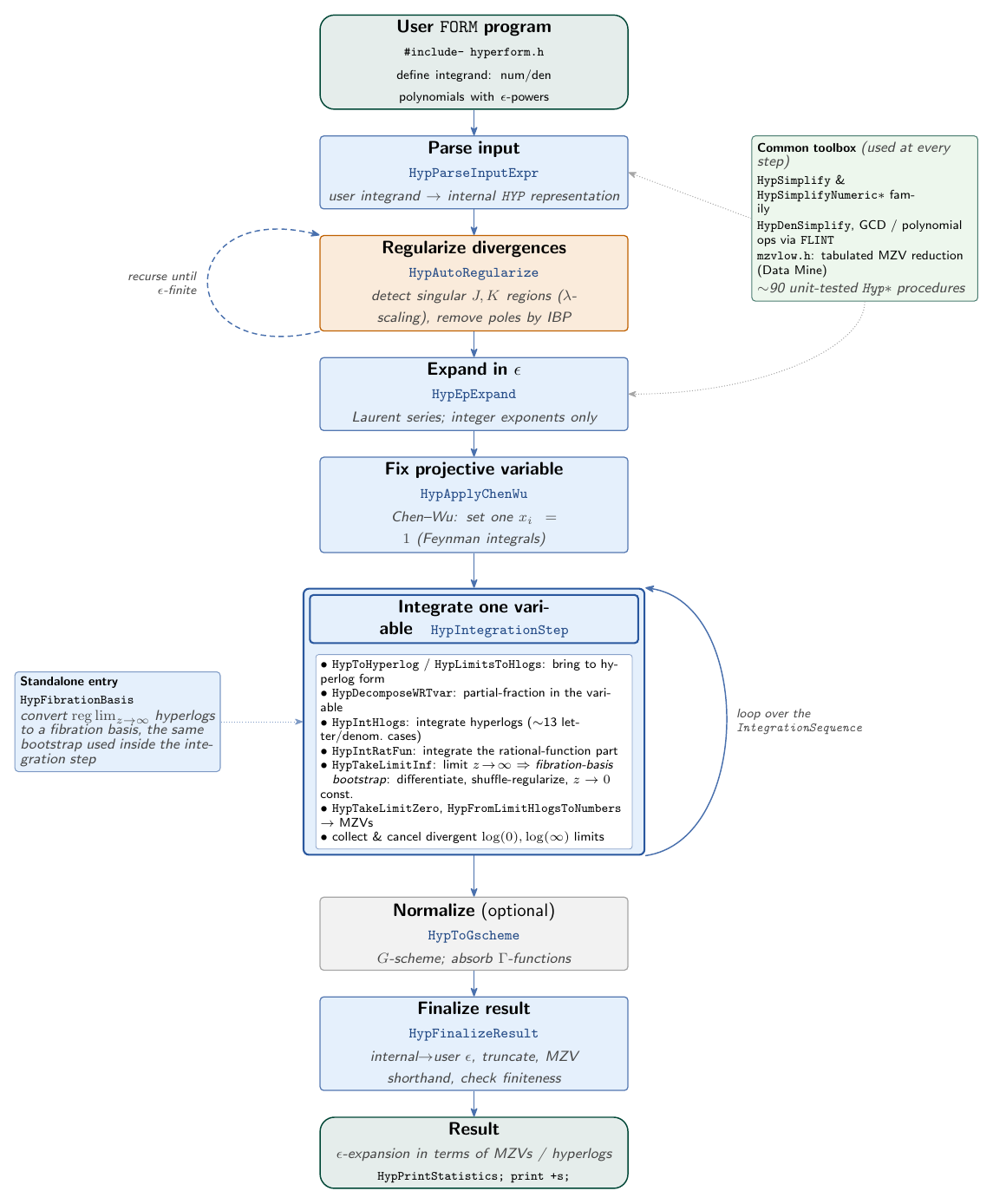}
  \caption{High-level workflow of the \ourcode{} package, showing the top-level
    procedures (typewriter font) and the internal sub-steps of the per-variable
    integration step. The common toolbox (simplification, polynomial and
    GCD operations via \flint{}, the tabulated MZV reduction) underlies every
    step; each of the $\sim$90 procedures is covered by dedicated unit tests.}
  \label{fig:flow}
\end{figure}

The overall workflow of a \ourcode{} calculation is summarized in
Fig.~\ref{fig:flow}. Starting from a user \form{} program, the integrand is
parsed into the internal representation, optionally regularized (recursively,
until $\ep$-finite), expanded in $\ep$ and reduced with the Chen--Wu gauge choice.
The heart of the package is the integration step \texttt{HypIntegrationStep},
applied once per variable in the chosen integration sequence: it converts the
integrand to hyperlogarithms, partial-fractions in the current variable,
integrates the hyperlogarithmic and rational parts, and takes the limits at
infinity and zero; the limit at infinity triggers the fibration-basis
``bootstrap'' (differentiation, shuffle regularization and a zero-limit
constant). A final normalization and finalization produce the result in terms of
MZVs. The same bootstrap is also exposed as a standalone entry point,
\texttt{HypFibrationBasis}.

\section{A complete example: the three-loop zigzag period}
\label{sec:z3}
To show how compact a full calculation is, we give the complete, self-contained
\form{} program for the three-loop zigzag period $Z_3$. It is an $\ep$-finite,
projective integral over six Feynman parameters, whose integrand is
$\delta(x_i-1)/\mathcal{U}_{Z_3}^2$ with the first Symanzik polynomial
$\mathcal{U}_{Z_3}$ (16 monomials). We fix $x_6$ via Chen--Wu and integrate
$x_1$ through $x_5$:
\begin{lstlisting}[style=form]
#include- hyperform.h
#define IntegralExpr "Z3"
#define IntegrationSequence "1,2,3,4,5"
#define ChenWuVar "6"
symbols x1,...,x6;
symbol ep;
cfunctions num,den;
local `IntegralExpr' =
  den(  x4*x5*x6 + x2*x5*x6 + x1*x5*x6 + x3*x4*x6 + x1*x4*x6 + x2*x3*x6
      + x1*x3*x6 + x1*x2*x6 + x3*x4*x5 + x2*x4*x5 + x2*x3*x5 + x1*x3*x5
      + x1*x2*x5 + x2*x3*x4 + x1*x3*x4 + x1*x2*x4 )^2;
.sort
#call HypParseInputExpr(ep,num,den,x1,...,x6)
.sort
#call HypEpExpand
#call HypApplyChenWu(`IntegralExpr',ChenWuVar)
#do IntVar={`IntegrationSequence'}
  #call HypIntegrationStep(`IntegralExpr',`IntVar')
#enddo
#call HypFinalizeResult(ep,`HYPMAXEP')
print +s;
.end
\end{lstlisting}
The program runs in about $0.05$\,s and returns the known result
\begin{equation}
  Z_3 = 6\,\zeta_3\,.
\end{equation}
No additional code is required; the listing above performs the entire
calculation: parsing, Chen--Wu gauge fixing, five integrations and finalization.

\section{Benchmarks: the zigzag family}
\label{sec:zigzag}
\begin{figure}[ht]
  \centering
  \begin{tikzpicture}[vertex/.style={circle,inner sep=0pt,minimum size=1.4mm,fill},
                      node distance=0.6cm]
    \node (1) [vertex] {};
    \node (3) [below right=of 1, vertex] {};
    \node (2) [above right=of 3, vertex] {};
    \node (4) [below right=of 2, vertex] {};
    \node (5) [above right=of 4, vertex] {};
    \node (6) [below right=of 5, vertex] {};
    \node (7) [above right=of 6, vertex] {};
    \draw (1)--(2); \draw (1)--(3); \draw (2)--(3); \draw (2)--(4);
    \draw (2)--(5); \draw (3)--(4); \draw (4)--(5); \draw (4)--(6);
    \draw (5)--(6); \draw (5)--(7); \draw (6)--(7);
    \draw (7) to[bend left=90,looseness=1.0] (1);
  \end{tikzpicture}
  \caption{The six-loop zigzag diagram $Z_6$. The long diagonal is fixed by the
    Chen--Wu gauge choice.}
  \label{fig:zigzag}
\end{figure}
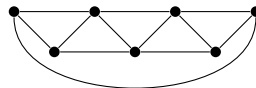

The zigzag graphs (Fig.~\ref{fig:zigzag}) make an especially convenient testing
ground. Their periods are finite in $\ep$ and linearly
reducible~\cite{brown2010periodsfeynmanintegrals}; carrying no external scales,
their second Symanzik polynomial drops out entirely, and closed forms are
available at every loop order. First conjectured by Broadhurst and
Kreimer~\cite{Broadhurst:1995km}, the values were later established by Brown and
Schnetz~\cite{Brown:2015ztw}, with a separate proof given in
Ref.~\cite{Derkachev:2023bpz},
\begin{equation}\label{eq:zigzagvalue}
  Z_n = 4\,\frac{(2n-2)!}{n!\,(n-1)!}
  \left(1-\frac{1-(-1)^n}{2^{\,2n-3}}\right)\zeta(2n-3)\,.
\end{equation}
In parametric form the period reads
$I_{Z_n}=\int_0^\infty\ud x_1\cdots\ud x_{2n}\,\delta(x_i-1)/\mathcal{U}_{Z_n}^2$,
with one variable fixed by the projective constraint.

Although finite and analytically known, the zigzags carry substantial
polynomial complexity, which makes them ideal for stress-testing the code.
Table~\ref{tab:zigzag} lists, per order, the Feynman-parameter count and the
number of monomials in $\mathcal{U}_{Z_n}$ (from the \form{} package
\texttt{uf}~\cite{adamkardos_2025_17646663}), with runtimes for \hyperint{},
\ourcode{} and \texttt{HyperlogProcedures}; \ourcode{} reaches $Z_6$, about $3.5$
times faster than \hyperint{} there. The \texttt{HyperlogProcedures} column
should be read with care, however: it rests on the theory of graphical
functions, an altogether different and far quicker route that only applies in
narrow settings
(massless, at most three external legs). What parametric integration offers in
return is generality.

\begin{table}[ht]
  \centering
  \setlength{\tabcolsep}{4.5pt}
  \begin{tabular}{c|c|c|c|c|c}
    Diagram & Variables & Monomials in $\mathcal{U}$ &
      \hyperint{}\,[s] & \ourcode{}\,[s] & \texttt{HyperlogProc.}$^\star$\,[s]\\
    \hline
    $Z_3$ &  6 &    16 & $0.8$ & $0.05$ & $0.04$ \\
    $Z_4$ &  8 &    45 & $1.5$ & $0.3$  & $0.04$ \\
    $Z_5$ & 10 &   130 & $54$  & $17$   & $0.04$ \\
    $Z_6$ & 12 &   368 & $1.0\times10^{5}$ & $2.9\times10^{4}$ & $0.05$ \\
    $Z_7$ & 14 &  1040 & --    & --     & $0.08$ \\
    $Z_8$ & 16 &  2919 & --    & --     & $0.13$ \\
  \end{tabular}
  \caption{Size of the zigzag problems, given by the number of Feynman parameters
    they carry and the monomials in their first Symanzik polynomial, with
    single-core runtimes from three programs. $^\star$\texttt{HyperlogProcedures}
    follows the graphical-functions approach, which is methodologically
    distinct.}
  \label{tab:zigzag}
\end{table}

\paragraph{Good integration sequences.}
Only some integration orders make the zigzags linearly reducible. Following
Brown~\cite{brown2010periodsfeynmanintegrals}, a good order completes as many
cycles as possible in the already-integrated subgraph while sharing as few
vertices as possible with the remainder; the largest such count is the
\emph{vertex width}. A graph of vertex width $\le 3$ is guaranteed linearly
reducible, and the zigzags admit an explicit width-three ordering, which is the
one used to obtain the \ourcode{} timings in Table~\ref{tab:zigzag}.

\begin{figure}[ht]
  \centering
  \begin{minipage}{0.45\textwidth}
    \centering
    \includegraphics[width=\linewidth]{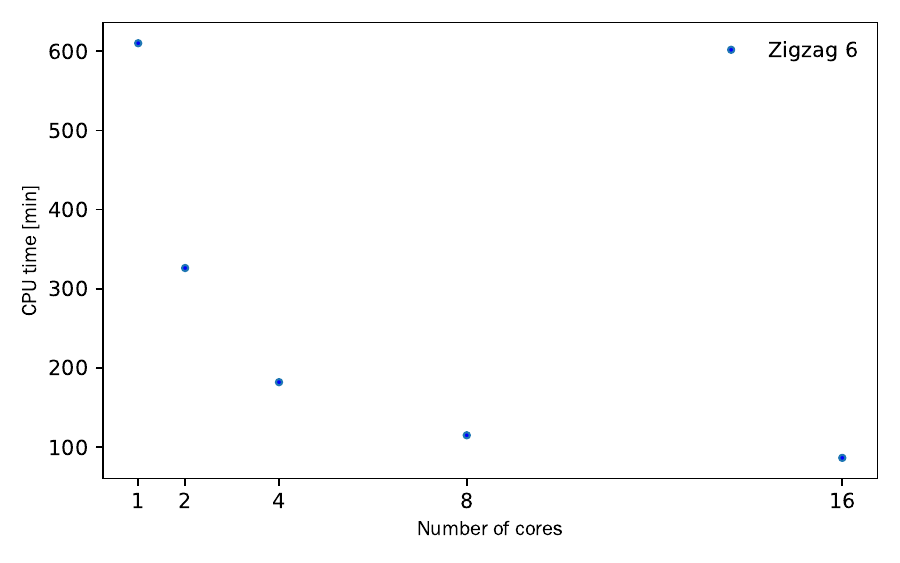}
    \caption{CPU time for $Z_6$ as a function of the number of \form{} workers.}
    \label{fig:workers}
  \end{minipage}
  \hfill
  \begin{minipage}{0.45\textwidth}
    \centering
    \includegraphics[width=\linewidth]{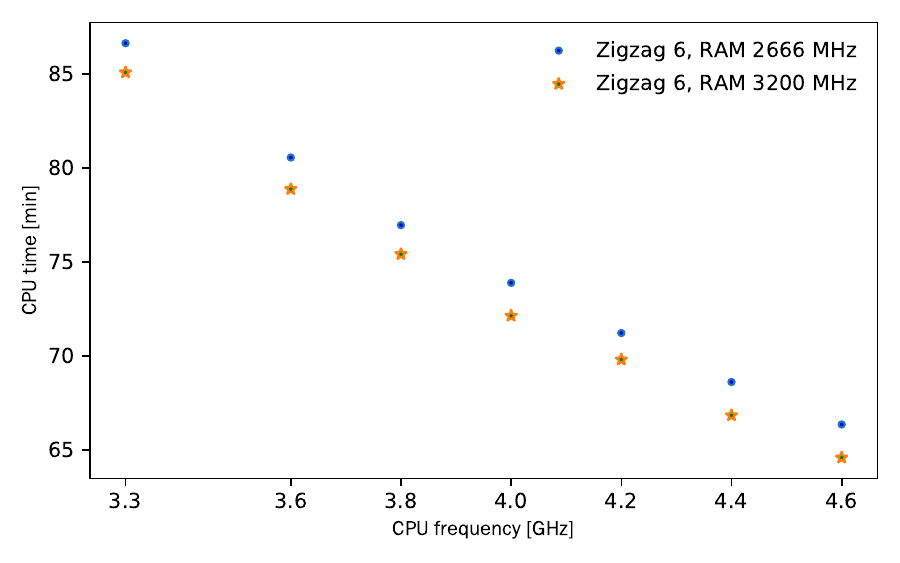}
    \caption{Computation time for $Z_6$ versus CPU clock frequency, for two RAM
      frequencies, using $16$ cores.}
    \label{fig:cpufreq}
  \end{minipage}
\end{figure}

\paragraph{Multicore scaling.}
\form{} runs on multicore, multi-CPU machines, yet parallelizing symbolic
manipulation is hard and a speed-up linear in the core count is essentially never
reached. Running $Z_6$ on an AMD~Ryzen~9~5900XT (16 physical cores, $3.3$\,GHz,
$128$\,GB RAM) with one to sixteen \form{} workers (Fig.~\ref{fig:workers}), the
CPU time falls substantially but sub-linearly and saturates, reflecting the
overhead of distributing and collecting terms. Overclocking the same machine
(Fig.~\ref{fig:cpufreq}), the CPU clock yields a nearly linear speed-up while the
RAM frequency matters far less; the multilayered cache keeps the cores well fed.

\section{Conclusions and outlook}
\label{sec:conclusions}
To summarize, \ourcode{} realizes hyperlogarithmic parametric integration, with
rational prefactors allowed, inside \form{}. It ports the \hyperint{} feature set
to a faster host and supplies, as a modular, unit-tested library of some ninety
procedures, the complete algorithmic chain: regularization in $\ep$, passage to a
fibration basis, the integration itself, regularized limits and MZV
reduction. On the zigzag suite it runs about $3.5$ times faster than \hyperint{}
on a single core, with \form{}'s multicore execution adding further gains.

This is a first release, with several improvements in view. Because polynomials
are at present stored explicitly, term sizes are bounded, which caps the highest
loop order reached at $Z_6$ for now; reworking this representation will
remove the ceiling. Beyond that we plan to suppress spurious polynomials via the Fubini
restriction and, more ambitiously on the mathematical side, to admit integrands
that go past strict linear reducibility. The code is openly
available~\cite{kardos_2025_17706909}, with the algorithms and a wider selection
of examples documented in the full paper.

\section*{Acknowledgments}
It is a pleasure to thank Francis Brown and Erik Panzer for numerous stimulating
exchanges, and Jos Vermaseren together with the entire \form{} development team
for the tools that made this work possible. The research was supported in part by
the ERC Advanced Grant 101095857 \emph{Conformal-EIC} (A.K.\ and S.M.) and by DFG
grant SCHN~1240/3-1 (O.S.). A.K.\ further thanks the EIC Theory Institute at
Brookhaven National Laboratory, where part of this work was carried out, and
acknowledges support from the University of Debrecen Program for Scientific
Publication.


\end{document}